\documentclass{article}
\usepackage{PRIMEarxiv}
\usepackage[utf8]{inputenc} 
\usepackage[T1]{fontenc}    
\usepackage{hyperref}       
\usepackage{url}            
\usepackage{booktabs}       
\usepackage{amsfonts}       
\usepackage{nicefrac}       
\usepackage{microtype}      
\usepackage{lipsum}
\usepackage{fancyhdr}       
\usepackage{graphicx}       
\graphicspath{{media/}}     
\usepackage{float}
\usepackage{longtable}
\usepackage{parskip}
\usepackage{tikz,graphics,color,fullpage,float,epsf,caption,subcaption}

\title{A Quantum Computing-driven Aid for New Material Design
}

\author{
  Kenson Wesley R \\
  School of Electronics Engineering \\
  Vellore Institute of Technology \\
  Chennai, India\\
  \texttt{kensonwesley.r2019@vitstudent.ac.in} \\
  \And
  Dr. Reena Monica P  \\
  School of Electronics Engineering \\
  Vellore Institute of Technology \\
  Chennai, India\\
  \texttt{reenamonica@vit.ac.in} \\
}

\begin{document}
\maketitle

\begin{abstract}
Material discovery is a phenomenon practiced since the evolution of the world.  The Discovery of materials had led to significant development in varied fields such as Science, Engineering and Technology etc., It had been a slow and long-drawn process, however, technological advancement had led to the rapid discovery of materials and the creation of a database that documented the earlier research and development.  Many intervening technologies at varying levels of efficiency were developed to advance the discovery of materials in the past and create a database.  Quantum computing is a recent development that further advances precision and accuracy.  In this study, the ground state energy of molecules such as GeO2, SiO2, SiGe, ZrO2 and LiH were found using the quantum algorithm Variational Quantum Eigensolver (VQE).  Also, a database consisting of the elements and molecules with the data of their Hamiltonian and Ground State energy was developed.
\end{abstract}


\section{Introduction}

Material discovery is a critical phenomenon practiced ever since human activity is recorded.   This has led to significant advancements in science, technology, medicine, engineering etc.,.  with wide-ranging applications.  For example, it has enabled the development of new materials with unique properties that have opened up new possibilities across many industries, such as electronics, energy, aerospace, Information Technology, healthcare etc.,  In the silicon industry, material discovery had led to the development of advanced materials with unique properties, such as graphene and the discovery of high-k dielectric materials, enabling the production of smaller and faster transistors for use in electronic devices.  In another development, the discovery of newer drugs with biocompatibility and delivery capabilities facilitated patient outcomes at a reduced cost.  However, each stage of development of materials or drugs is an advancement towards precision and accuracy that facilitates robust and informed decision-making.  Of late, this has been aided by quantum computing algorithms.  This study was aimed at finding the ground state energy of a few molecules such as GeO2, SiO2, SiGe, ZrO2 and LiH using Variational Quantum Eigensolver (VQE).  It also aims at the creation of a database \cite{12} of elements and molecules with the data of their Hamiltonian and Ground state energy.  This database enables researchers and practitioners in the discovery and designing of materials with desired properties.
\vspace{5mm}

\section{Literature Review}

Machine learning and Artificial Intelligence (AI) are the most debated topics in the last few years.  Of late, the focus shifted to another connected area, quantum computing.  Quantum computing is becoming a reality with the technology world's increasing implementation of qubits. This was augmented by cloud computing which has made researchers and those in practice remotely use quantum computers for their purpose.  Jaeho et al., \cite{4} reported that useful quantum computing techniques were introduced for AI engineers. These techniques have the potential to solve complex optimization problems, improve machine learning algorithms, and enhance AI applications. As the world moves towards making quantum computing techniques more feasible, a new realm of possibilities for numerous AI engineers emerges.

An optimised cost function was put forth by Han et al. \cite{1} for the meta-VQE technique, which is utilised to determine ground state energy on NISQ devices. Two more indications are included in the revised cost function to increase accuracy and increase the penalty degree of inaccuracy. The approach is used in conjunction with quantum architecture search to get over the fixed structure restriction. Results demonstrate that the improved method can more accurately estimate the length of the equilibrium bond by learning characteristics of ground state energy changes in noisy channels.

In order to lower the computing cost of the Variational Quantum Eigensolver (VQE) algorithm, Pranav et al., \cite{3} developed a novel method that takes advantage of the simultaneous measurability of subproblems that correspond to the commuting terms. The method transforms VQE instances into a pattern designed for simultaneous measurement, which results in an 8–30x lower cost. By employing a quantum computer to estimate the deuteron's ground state energy, the method is experimentally verified. To mitigate undesirable covariance terms, an adaptive strategy is developed.

Julian et al., \cite{2} used Gaussian Processes (GP) to reduce noise and provides high-quality seeds to escape local minima. The method outperforms ImFil on larger dimensional problems, but not consistently for multimodal landscapes.

VQE’s scalability is a significant hindrance to its practical implementation, as it grows in complexity with the fourth power of the molecule's size. However, researchers have found that if the components are grouped into smaller, compatible families, the algorithm's complexity can be reduced to the third power. This has been demonstrated through a process developed by Pranav et al. \cite{5}, which partitions the components and creates the necessary quantum circuits. This approach has been experimentally validated through the estimation of the deuteron's ground state energy using a 20-qubit IBM machine.

J.R. et al., \cite{6} described a new method using the two-photon "R-transfer" scheme is developed to produce translationally cold potassium molecules in the ground state. Molecules are detected using two-color resonant ionization with pulsed lasers, and the translational temperature is measured to be approximately 500 $\mu$K.

To investigate how well quantum probes can be identified and to put quantum identification techniques into practise, quantum Hamiltonian identifiability is essential. In order to apply it to the quantum Hamiltonian identification domain, Yuanlong Wang et al. \cite{7} generalise the identifiability test based on the similarity transformation method (STA) in classical control theory. With the use of single-qubit probes, the STA is utilised to establish the identifiability of spin-1/2 chain systems with arbitrary dimensions. For nonminimal systems, the structure preserving transformation (SPT) approach is put forth, adding a sign that practical quantum Hamiltonian identification procedures exist. Finally, a Hamiltonian identification algorithm for the economy is described with an example, and simulations are used to show how successful it is.

In order to determine the ground-state energy of sodium hydride using the variational quantum eigensolver technique, Jerimiah et al. \cite{9} looked into the impact of device noise on the precision and fidelity of quantum circuits. They put into practise a number of ansatz circuits derived from unitary coupled cluster theory and investigated how the errors brought about by gate-based noise, the depth of the ansatz circuit, parameter optimisation techniques, and internal nuclear configuration impacted the relative error in the energy and fidelity of the prepared quantum states. Insights from the findings can be used to create effective ansatz circuits that can manage noisy quantum hardware in applications for near-term quantum computing.

A precise analytical estimate of the ground state electron density of neutral atoms from He to Lr is reported. The approximation is stated as a linear combination of up to 30 basis functions, and it possesses traits like non-negativity, normalisation, Hartree-Fock moment replication, adherence to the cusp condition, and accurate exponential decay in the long-range asymptotic area. For many uses in physics and chemistry, this approximation, denoted F(r), reproduces the Hartree-Fock electron density to first order in perturbation theory. \cite{11}
\vspace{5mm}

\section{Methodology}
\vspace{2mm}
\subsection{Algorithm Used}
Using the VQE algorithm to simulate the properties of different chemical compounds involves a series of steps. Firstly, the molecular structure of the compound must be defined, including the number of atoms, their positions, and the type of bonds between them. A suitable quantum simulator, either cloud-based or local, must then be chosen, along with an appropriate basis set. The VQE algorithm can then be implemented using a quantum programming language, involving the construction of an ansatz and a Hamiltonian. The ansatz parameters are optimized using classical optimization algorithms, and the properties of the molecule, such as its ground-state energy and dipole moment, are calculated. These results can be compared with experimental data or other computational methods to assess the accuracy of the simulation. The process can be repeated for different compounds, varying the molecular structure and optimizing the ansatz for each individual compound. The accuracy of the simulation can be improved by using a larger basis set or a more complex ansatz, although this may require more computational resources.

\subsection{Hartree-Fock ground state}

The Hartree-Fock method is a popular method for approximating the ground state electronic structure of molecules and compounds. We define the molecular structure of the system of interest, including the positions of all atoms. Using electronic Hamiltonian, describes the energy of the electrons in the system. This involves computing the overlap integrals, kinetic energy integrals, and Coulomb and exchange integrals for the molecular orbitals.
We then perform the Hartree-Fock self-consistent field (SCF) procedure, which involves iteratively calculating the molecular orbitals and the electronic energy until self-consistency is reached. This is typically done using an algorithm such as the Roothaan-Hall or direct inversion of the iterative subspace (DIIS) method.
We calculated the overall energy of the Hartree-Fock ground state once self-consistency was achieved by adding the energies of the occupied molecular orbitals. We can additionally calculate other Hartree-Fock ground state characteristics, including the molecule's polarizability or dipole moment.

\subsection{Electronic Hamiltonian}

The Schrödinger equation states: H$\Psi$ = E$\Psi$, where H is the Hamiltonian operator, $\Psi$ is the system's wave function, and E is the total energy of the system. The electronic Hamiltonian defines the total energy of a system of interacting electrons and nuclei. The kinetic energy terms for the electrons and nuclei, potential energy terms resulting from their interaction, and any additional potential energy terms make up the Hamiltonian operator.
Determine the atomic or molecular structure of the system. This involves identifying the positions of the nuclei in the system and the number and positions of the electrons. The structure can be determined experimentally or computationally using molecular modeling methods such as density functional theory (DFT) or Hartree-Fock theory. Once the atomic or molecular structure has been determined, the next step is to choose a suitable theoretical approach to calculate the electronic Hamiltonian. This can be done using methods such as DFT or Hartree-Fock theory. The choice of method will depend on the specific system being studied and the level of accuracy required. The electronic Hamiltonian can then be solved using the chosen theoretical approach. This involves solving the Schrödinger equation using techniques such as variational methods or matrix diagonalization.
\vspace{5mm}
\section{Results}
\vspace{2mm}
\subsection{VQE}

Using VQE, we’re able to find the energy of the molecule with varying atomic distance from 0.3 to 4.0. We’re are also able to compare the exact energy vs the energy calculated by VQE. We take the results for molecules used in different areas of field. GeO2 is used as a high-k dielectric material for gate insulation in metal-oxide-semiconductor (MOS) devices. We have also taken SiO2 which is used in used conventionally for the same purpose as GeO2 but has lower dielectric constant. SiGe is in the fabrication of high-speed and low-power electronic devices, such as field-effect transistors (FETs) and radio-frequency (RF) devices. ZrO2, also known as zirconia or zirconium dioxide, is a ceramic material that is widely used in various industrial and biomedical applications. LiH is an important material in the field of energy storage and conversion. It is used as a hydrogen storage material for fuel cells and other hydrogen-based technologies, due to its high hydrogen content (7.8 wt\%) and low molecular weight.

\begin{figure}[H]
  \centering
  \begin{minipage}[b]{0.47\textwidth}
    \includegraphics[width=\textwidth]{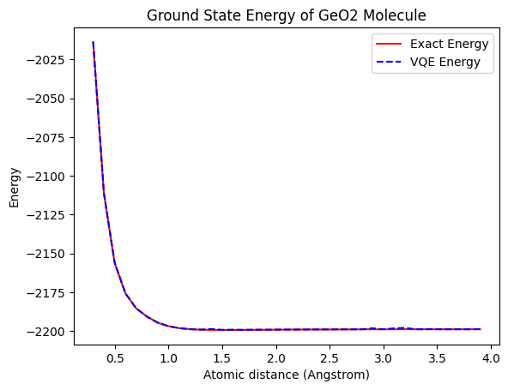}
    \caption{VQE energy for GeO2}
  \end{minipage}
  \hfill
  \begin{minipage}[b]{0.47\textwidth}
    \includegraphics[width=\textwidth]{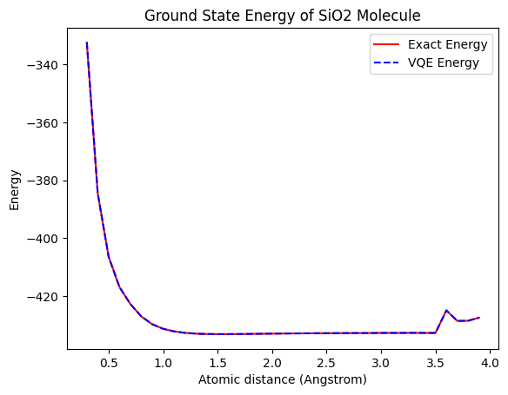}
    \caption{VQE energy for SiO2}
  \end{minipage}
\end{figure}

\begin{figure}[H]
  \centering
  \begin{minipage}[b]{0.47\textwidth}
    \includegraphics[width=\textwidth]{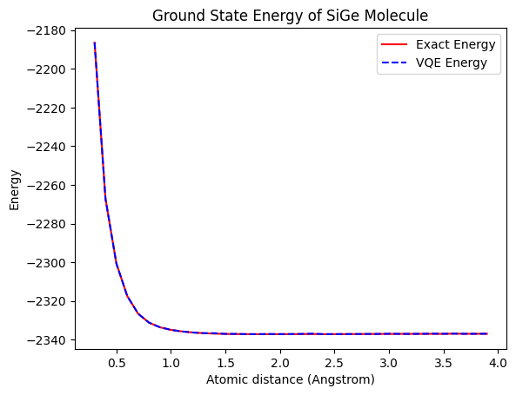}
    \caption{VQE energy for SiGe}
  \end{minipage}
  \hfill
  \begin{minipage}[b]{0.47\textwidth}
    \includegraphics[width=\textwidth]{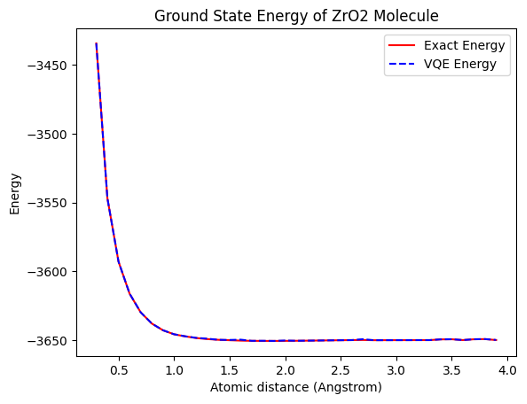}
    \caption{VQE energy for ZrO2}
  \end{minipage}
\end{figure}

\begin{figure}[H]
    \centering
    \includegraphics[height=75mm,scale=0.25]{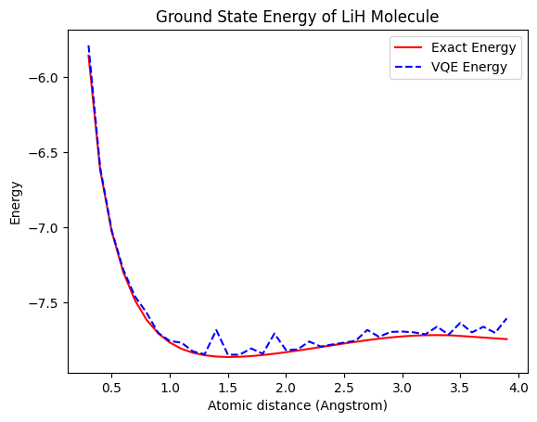}
    \caption{VQE energy for LiH}
    \label{fig:LiH}
\end{figure}
\vspace{5mm}
\subsection{Hartree-Fock Ground state energy }

\begin{figure}[H]
    \centering
    \includegraphics[height=75mm,scale=0.25]{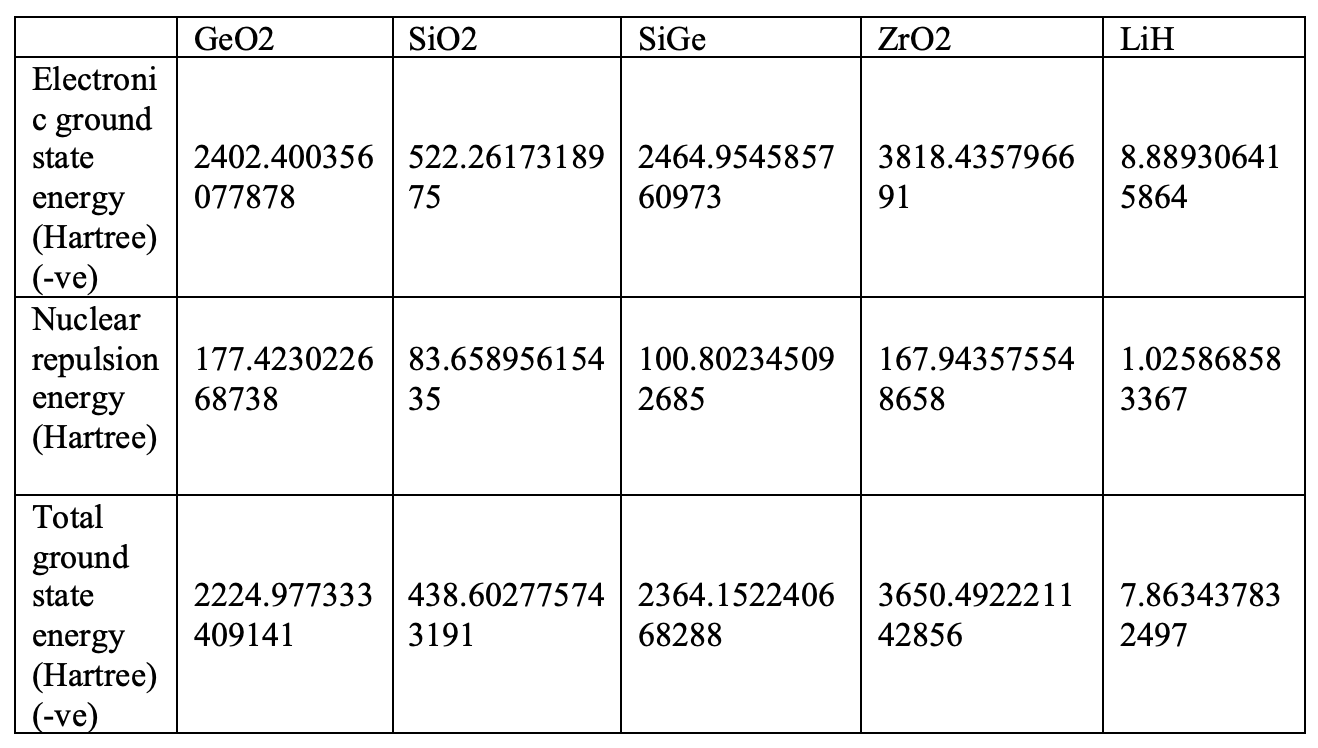}
    \caption{Total ground state energy for GeO2, SiO2, SiGe, ZrO2 and LiH}
    \label{Tab:Hartree}
\end{figure}

\vspace{2mm}
\subsection{Hamiltonian using Hartree-Fock method}
\vspace{2mm}
For a C2 molecule, the number of spin orbitals we get is 20 and the number of terms in total is 8508; the full Hamiltonian for C2 and other molecules is given in the database. These are the first fifteen terms:

$-19.945046536186272 * ( +_0 -_0 ) \\
+ 2.4294342059905365e-08 * ( +_0 -_2 ) \\
+ -0.4382689806514389 * ( +_0 -_3 ) \\
+ 1.0736319223868062e-08 * ( +_0 -_6 ) \\
+ -0.1692105628741365 * ( +_0 -_9 ) \\
+ -19.9461447511049 * ( +_1 -_1 ) \\
+ 0.40909845936737327 * ( +_1 -_2 )\\
+ 2.602662301936696e-08 * ( +_1 -_3 )\\
+ 0.18079318929582455 * ( +_1 -_6 )\\
+ 1.004855655874829e-08 * ( +_1 -_9 )\\
+ 2.4294342127863153e-08 * ( +_2 -_0 )\\
+ 0.40909845936737366 * ( +_2 -_1 )\\
+ -6.533602041278337 * ( +_2 -_2 )\\
+ 0.5644788196496222 * ( +_2 -_6 )\\
+ -0.438268980651439 * ( +_3 -_0 )\\ $
\vspace{5mm}
\section{Discussion}
\vspace{2mm}

With the help of VQE we have taken the ground state of few molecules in different bond length. We can see how the total energy drops as the distance increases drops but increases in the end for LiH and SiO2. We have taken the Hartree-Fock ground state which falls in line with the calculated ground state using VQE. Since the Electronic ground state is the pure energy, we need to find the nuclear repulsion energy also and after adding both the energies we get the total Hartree-Fock ground state of each molecule. The Hamiltonian was found after giving the 2-body terms and using Hartree-Fock method. The calculated ground state energy and the Hamiltonian for various molecules and elements and have been keyed into the database created. With this database, one can predict energy levels which we have done with VQE for few molecules, predict their properties like electron density distribution or its magnetic moment and optimize that said properties. With these we can design new materials with desired properties. We can use the Hamiltonian to calculate the desired property for different atomic configurations. For example, we can calculate the thermal conductivity of different lattice structures and atomic spacings, and compare the results. 
\vspace{5mm}
\section{Conclusion and Future works}
\vspace{2mm}
In conclusion, this paper presents the calculated ground state energy of GeO2, SiGe, SiO2, ZrO2 and LiH using VQE and have compared it to the exact energy of each molecule respectively. We have also found the Hartree-Fock ground state energy for each of the molecules and have created a database for many elements and molecules with their Hamiltonian and the Hartree-Fock ground state energy. Heaving elements and compunds can be added to the database and the accuracy can be improved to for very precise values. VQE can be run to find the ground state for new compounds with the experimental data.

\end{document}